\documentclass[runningheads]{llncs}
\usepackage[T1]{fontenc}
\let\llncssubparagraph\subparagraph
\let\subparagraph\paragraph
\usepackage[compact]{titlesec}
\let\subparagraph\llncssubparagraph
\usepackage{titlesec}
\titlespacing*{\section}{0pt}{2ex}{2ex}

\usepackage{booktabs}
\usepackage{array}
\usepackage{lipsum}
\usepackage{xcolor}
\usepackage{fancyhdr}
\usepackage{tikz}
\usepackage{pifont}
\usepackage{cite}
\usepackage{url}
\usepackage{graphicx} 
\usepackage{subcaption}
\usepackage[hidelinks,colorlinks=true,linkcolor=blue,citecolor=blue]{hyperref}
\usepackage{subcaption}
\usepackage{multirow}
\usepackage{newtxtext}       %
\usepackage{newtxmath} 
\usepackage{graphicx}
\usepackage{orcidlink}
\newcolumntype{C}[1]{>{\centering\arraybackslash}p{#1}}%
\newcommand{\circled}[1]{%
  \begin{tikzpicture}[baseline=(char.base)]
    \node[draw, circle, fill=black, text=white, inner sep=0.6pt] (char) {\textbf{#1}};
  \end{tikzpicture}
}
\usepackage{color}
\usepackage{hyperref}
\hypersetup{
    colorlinks=true,      
    linkcolor=red,       
    anchorcolor=black,    
    citecolor=blue,      
    filecolor=cyan,       
    menucolor=red,        
    runcolor=cyan,        
    urlcolor=magenta      
}

\begin{document}
\title{Malware Classification Leveraging NLP \& Machine Learning for Enhanced Accuracy}
%
%
\author{Bishwajit Prasad Gond\inst{1}\orcidlink{0000-0003-3640-0463}\and
Rajneekant\inst{1} \and
Pushkar Kishore\inst{1}\and
Durga Prasad Mohapatra\inst{1}}
\authorrunning{B. P. Gond et al.}
%
\institute{National Institute of Technology Rourkela, Odisha, India\\
\email{bishwajitprasadgond@gmail.com}, \email{rajneekant200@gmail.com},\\ \email{monumit46@gmail.com}, and \email{durga@nitrkl.ac.in}}
\maketitle              
\begin{abstract} This paper investigates the application of natural language processing (NLP)-based $n$-gram analysis and machine learning techniques to enhance malware classification. We explore how NLP can be used to extract and analyze textual features from malware samples through $n$-grams, contiguous string or API call sequences. This approach effectively captures distinctive linguistic patterns among malware and benign families, enabling finer-grained classification. We delve into $n$-gram size selection, feature representation, and classification algorithms. While evaluating our proposed method on real-world malware samples, we observe significantly improved accuracy compared to the traditional methods. By implementing our $n$-gram approach, we achieved an accuracy of 99.02\% across various machine learning algorithms by using hybrid feature selection technique to address high dimensionality. Hybrid feature selection technique reduces the feature set to only 1.6\% of the original features.
\end{abstract}
\section{Introduction}\label{sec:intro}
In the relentless cat-and-mouse game of cybersecurity, where threats perpetually adapt and multiply, it has become increasingly crucial to deploy innovative techniques for malware detection and classification. Malwares, spanning a wide spectrum of categories such as adware, viruses, worms, spyware, downloaders, trojans, and backdoors, pose an escalating challenge to security experts worldwide. To combat this diverse landscape of threats, the incorporation of Natural Language Processing (NLP)-based $n$-gram\cite{moskovitch2008unknown} analysis, particularly $n$-grams based feature selection, has emerged as a promising avenue for enhancing malware classification. This paper explores the application of $n$-grams analysis, coupled with machine learning, to classify malwares into these above seven distinct categories.

Beyond traditional signature-based \cite{bergeron2001static} and heuristic-based approaches\cite{bazrafshan2013survey}, which often struggle to cope with polymorphic and obfuscated malware\cite{alrzini2020review}, our methodology leverages the unique linguistic footprints exhibited by different malware categories. Specifically, we focus on the text extracted from the Application Programming Interface (API) calls of Portable Executable (PE) files in Windows environments\cite{li2022dmalnet}. 

Our research focuses on fine-grained malware classification by leveraging $n$-grams and machine learning techniques to identify intricate patterns in API call sequences of different malware variants. This novel approach promises accurate and nuanced classification, providing a robust defense against evolving threats.

Coupling $n$-grams with machine learning in malware classification is an inventive approach enhancing our ability to discern subtle behavioral and structural differences. Utilizing API call sequences in PE files broadens the scope beyond static analysis to dynamic behavior monitoring.

As the cybersecurity community seeks more effective and adaptive defenses, this research aims to contribute by exploring $n$-grams, machine learning techniques, and API call sequences for improved malware categorization accuracy and efficiency.

Dynamic analysis based on API calls tends to be more effective than static analysis \cite{raff2020survey}. This paper investigates using natural language processing (NLP) techniques for dynamic malware analysis based on $n$-gram API calls, leveraging ensemble and machine learning methods like boosting (XGBoost and LightGBM) and bagging (Random Forest) to enhance malware classification accuracy.

The rest of the paper is organized as follows: in Section \ref{sec:basicconcepts} we covered the basic concepts. In Section \ref{sec:relatedwork}, we review recent literature using API sequence and NLP in malware classification and detection. Section \ref{sec:framework} outlines our NLP-based framework for malware detection, focusing on preprocessing and the NLP process. Section \ref{sec:experiment} details our experimental setup, including obtaining an API key, data extraction and malware classification. Section \ref{sec:rana} covers the detail result analysis. In Section \ref{sec:compare}, we compare our work with state-of-the-art techniques in malware detection and classification. Finally, Section \ref{sec:futurework} concludes our work and outlines some future research directions.

\section{Basic Concepts}\label{sec:basicconcepts}
In this section, we first discuss the static analysis of malware, followed by dynamic analysis. We also address the challenges associated with these approaches. Furthermore, we delve into sandbox environments and explore current trends in malware analysis.
\subsection{Static Analysis}
Static analysis dissects malware's binary code, file structure, strings, and metadata without execution. It identifies functions, suspicious code, attack vectors, and indicators of compromise (IOCs), aiding rapid threat detection and classification. However, it has limitations in uncovering behavior details, especially with obfuscation techniques like code encryption, packing, anti-analysis measures, polymorphism, and control flow obfuscation. Security experts often combine static analysis with dynamic analysis and reverse engineering for comprehensive understanding.

\subsection{Dynamic Analysis}
Dynamic analysis involves executing malware in a controlled sandbox environment to observe real-time behavior, crucial for understanding capabilities and developing countermeasures. As malware sophistication grows with polymorphic and metamorphic coding, dynamic analysis gains prominence. Sandboxes simulate real environments, capturing behaviors like file system, registry, network activities, enabling comprehensive monitoring.

Current trends focus on automation and machine learning for efficient large-scale analysis and accurate pattern recognition. Automated tools rapidly process vast samples, identifying common malicious behaviors. Machine learning techniques enhance threat detection accuracy, adapting to evolving tactics.

\section{Related Works}\label{sec:relatedwork}
.
This section discusses research on malware detection and classification related to Natural Language Processing (NLP) and API Call.

Nakazato et al.\cite{nakazato2011novel}  proposed an approach to detect and classify malware in their malwares using API sequence. They conducted dynamic analysis to automatically obtain the execution traces of malware and then classified malwares into clusters based on their behavior characteristics derived from Windows API calls in parallel threads. The authors utilized two NLP techniques, namely $n$-gram and TF-IDF, to infer the characteristics of malware samples. This proposed methodology successfully classified 90\% of 2312 malware samples into maximum of 20 different clusters.

Another research on malware related to API sequences and machine learning was proposed by Chandrasekar et al.\cite{ravi2012malware}, who used a $3^{rd}$ order Markov chain to model Windows API call sequences. Their proposed malware detection system tested on a dataset containing 94 benign and 179 malware executables, achieved an accuracy of 90\%.

Windows API calls are highly deterministic features for behavior-based malware detection, accurately reflecting program behavior during execution and effectively distinguishing malware from benign programs \cite{han2019maldae,ye2009sbmds}. API call information can identify evasive malware. Hence, many malware detection techniques utilize methods for extracting and leveraging API call information.

Ye et al.\cite{ye2009sbmds} developed an intelligent malware detection system, IMDS, using API features. They extracted API calls accessed by PE files through static analysis and used these features as input for classification algorithms, achieving a detection accuracy of 93.07\%. Additionally, authors proposed a layered classification framework to determine malicious operations performed by malware samples using API call features. They applied various feature selection methods to obtain discriminative features, resulting in an accuracy of 98.6\%.

Dabas et al.\cite{dabas2023effective} introduced a malware detection method for Windows based on API calls, using three feature sets: (a) API calls usage, (b) frequency, and (c) sequences, enriched with TF-IDF to create the API integrated feature set. Malware samples are retrieved from the
VirusShare data repository having 2500 malware and 2500 benign samples from freshly installed Windows 10 OS. It achieves 99.6\% accuracy across ML algorithms, addressing high-dimensionality with hybrid feature selection, reducing feature set to 9\% .

Dabas et al.\cite{dabas2023malanalyser} presented MalAnalyser, a lightweight Windows malware detection system based on frequent API call subsequences. The authors collected malwares from VirusShare platform having 2500 malware and 2500 benign samples from freshly installed Windows 10 OS. It uses GLBPSO and GA for feature selection, achieving up to 99.7\% accuracy with 30\% features, and 100\% accuracy on GA-enriched features. MalAnalyser outperforms similar approaches on a benchmark dataset with 99.72\% accuracy.

Sharma et al.\cite{sharma2022windows} proposed an approch which paper focuses on Windows malware detection using TF-IDF enriched API call information. They combined API call feature sets into an integrated set and applied TF-IDF for feature importance and the dataset was obtained from Virusshare. The TF-IDF enriched integrated API calls feature set achieves up to 99.91\% accuracy for SVM and Logistic Regression algorithms.
\section{Proposed Framework}\label{sec:framework}
In this section, we present out proposed framework for malware classification using NLP and machine learning techniques.
\begin{figure*}[ht!]
  \centering
  \includegraphics [angle=0,origin=c,width=0.99\textwidth]{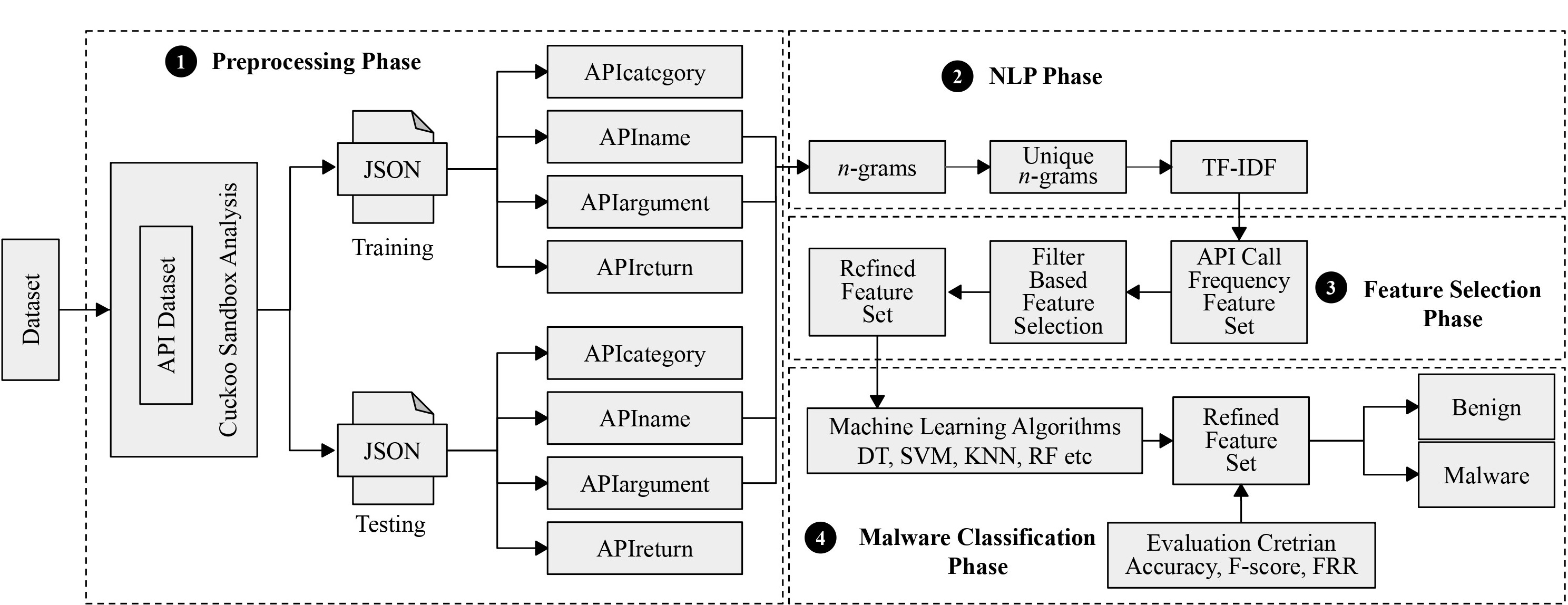}
  \caption{Proposed Architecture for Malware Classification }
  \label{fig:nlpsim}
\end{figure*}
Our proposed architecture (Fig. \ref{fig:nlpsim}) involves: \\ 1) acquiring malware hash from VirusShare\footnote{\url{https://virusshare.com}}, 2) querying VirusTotal\footnote{\url{https://www.virustotal.com}} for JSON file with antivirus scans to determine malware class, 3) downloading distinct malware categories, 4) dynamic analysis in Cuckoo sandbox\footnote{\url{https://cuckoosandbox.org}} to extract API call sequences in JSON format, 5) partitioning JSON report into API name, argument, return, and category text files, 6) applying $n$-gram methods combining API names and arguments, 7) calculating TF-IDF using unique $n$-grams from all categories, 8) applying hybrid feature selection techniques to get refine feature set, 9) applying machine learning techniques and adjusting evaluation criterion on refine feature set. The four phases are:

\noindent\textbf{\circled{1} Preprocessing Phase:} Steps 4 and 5.

\noindent\textbf{\circled{2} NLP Phase:} Step 6 and 7.

\noindent\textbf{\circled{3} Feature Selection Phase:} Step 8.

\noindent\textbf{\circled{4} Malware Classification Phase:} Step 9.

\subsection*{Phase 1: Preprocessing Phase}
In preprocessing phase, we perform the following (Fig. \ref{fig:dpfe}) activities:
\begin{figure*}[ht!]
  \centering
  \includegraphics [angle=0,origin=c,width=0.99\textwidth]{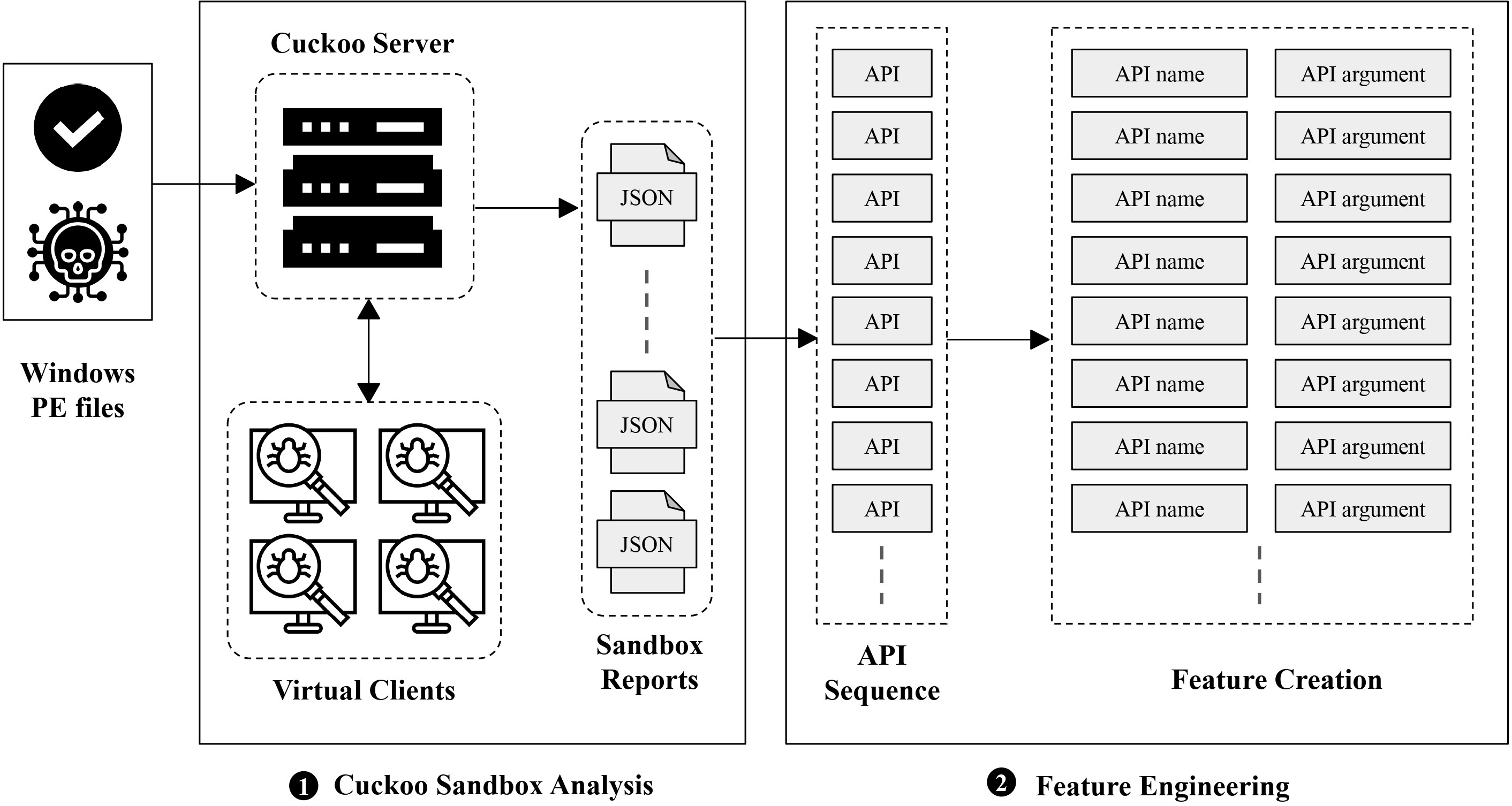}
  \caption{Data Preprocessing and Feature Engineering}
  \label{fig:dpfe}
\end{figure*}
\begin{enumerate}
\item \textbf{Preprocess the Data:}
After obtaining the data for each class, we performed behavioral analysis using Cuckoo Sandbox, resulting in a behavioral report in JSON format. We then split the file into four parts: API category, API name, API argument, and API return. From this, we selected the API name and argument to create n-grams, where the API name is the first part and the API call argument is added using underscores. Finally, we obtained a CSV file containing n-grams for each category.
    \item \textbf{API Dataset \& Cuckoo Sandbox Analysis:} The process starts with taking a dataset that undergoes Cuckoo Snadbox Analysis which results the behavioural report of Portable Executable in JSON format.
    \item \textbf{Data Division:} The data is divided into training and testing sets in 80\% and 20\%, both processed through JSON format.
    \item \textbf{API Elements Extraction:} Various API elements like APICategory, APIName, APIArgument, and APIreturn are extracted.
\end{enumerate}

\subsection*{Phase 2: NLP Phase}
In NLP phase, we perform the following activities:
\begin{enumerate}
    \item \textbf{$n$-grams and Unique $n$-grams:} We extract $n$-grams and select Unique  $n$-grams from the JSON.
    \item \textbf{$n$-grams after Processing:} The followings are the examples of  unigram, bigrams and trigrams that we have used in this paper.
\begin{itemize}
    \item \textbf{Unigram: }\textit{LdrLoadDll\_urlmon\_urlmon.dll}
    \item \textbf{Bigram: }\textit{NtAllocateVirtualMemory\_na,\\ LdrLoadDll\_ole32\_ole32.dll}
    \item \textbf{Trigram: }\textit{LdrUnloadDll\_SHELL32,\\ LdrLoadDll\_SETUPAPI\_SETUPAPI.dll,\\ LdrGetProcedureAddress\_ole32\_OleUninitialize}
\end{itemize}
    \item \textbf{Vectorization with TF-IDF:}
It tokenizes text, counts occurrences of each token, and computes TF-IDF weights, which reflects the importance of each token in a document relative to the entire corpus. Optional L2 normalization ensures consistent feature vector lengths. TF-IDF is applied on each csv files of unigram, bigram and trigram with Unique $n$-gram csv file to transform the text data.
    \item \textbf{Refined Feature Set:} A refined feature set is obtained after filtering the unnecessary feature of each $n$-grams CSV file.
\end{enumerate}
\subsection*{Phase 3: Feature Selection Phase}Out of the millions of features available, we have applied several hybrid feature selection and reduction methods to eliminate redundant features. These methods eliminate features containing numbers, special characters, or underscores, which are prevalent in both malware and benign classes within our feature set. The goal is to identify and retain only the most relevant features for classification. This process helps reduce the dimensionality of the dataset, which can improve the efficiency and effectiveness of machine learning algorithms. By selecting the most informative features and discarding irrelevant or redundant ones, we aim at enhancing the performance of the classification model.
In feature selection phase, we perform the following activities:

\begin{enumerate}
    \item \textbf{API Call Frequency Feature Set:}We explore the features derived from API call frequency to understand the system behavior and usage patterns.   
    \item \textbf{Filter Based Feature Selection:} 
    We apply filter-based techniques (e.g., mutual information, correlation analysis) to select the most informative features.
    \item \textbf{Refine Feature Set:} We eliminate the redundant or irrelevant features to ensure that the final set is discriminative and predictive.
\end{enumerate}

\subsection*{Phase 4: Malware Classification Phase}
In malware classification phase, we perform the following activities:
\begin{enumerate}
    \item \textbf{Applying Machine Learning Algorithms:} Algorithms such as Decision Tree \cite{lavarnway1988introduction}, SVM with Kernal: linear, sigmoid, polynomial $3^{\circ}$, polynomial $4^{\circ}$ and RBF, KNN, Random Forest \cite{breiman2001random}, XGBoost, LightGBM are applied to the refined feature set obtained from phase three.
    \item \textbf{Use Evaluation Criteria:} We use the evaluation criteria such as accuracy, precison, recall and $F_1$ score, etc. to determine the effectiveness of these ML algorithms.
    \item \textbf{Classify the Malwares :} This phase involves distinguishing between benign and malware entities based on the features selected and evaluated in the previous phases.
\end{enumerate}
To gain an in-depth understanding of the architecture, one can
 refer to the source code of our malware detector available on GitHub\footnote{\url{https://github.com/bishwajitprasadgond/MalwareClassification}}

Our research utilizes malware samples collected from VirusShare. We obtained an API key to download over 200,000 JSON files (each corresponding to a unique hash), but were limited to four files per minute due to API constraints. From these JSON files, we extracted essential information, including scan results from 70 antivirus programs, which served as the basis for subsequent malware classification. We then performed malware classification, a crucial step to identify and categorize the samples for further analysis.

\section{Working Example}
\noindent\textbf{Example Documents:}
\begin{itemize}
    \item A: "This is a sample text document."
    \item B: "Here is another text document."
\end{itemize}

\subsection*{TF-IDF Calculation}
\begin{enumerate}
    \item \textbf{Term Frequency (TF):} $TF(t, d) = \dfrac{f_{t,d}}{\sum_{t' \in d} f_{t',d}}$\\
        $f_{t,d}$ is the term frequency, $\sum_{t' \in d} f_{t',d}$ is total terms in $d$.
    \item \textbf{Inverse Document Frequency (IDF):} $IDF(t, D) = \log \left( \dfrac{N}{|\{d \in D : t \in d\}|} \right)$\\
        $N$ is total docs, $|\{d \in D : t \in d\}|$ is docs containing $t$.
    \item \textbf{TF-IDF:} $\text{TF-IDF}(t, d, D) = TF(t, d) \times IDF(t, D)$
\end{enumerate}

\noindent\textbf{Example:}
\begin{itemize}
    \item IDF("sample") = IDF("another") = IDF("text") = IDF("document") = 0.3
    \item TF("sample", A) = TF("sample", B) = 1/5 = 0.2
    \item TF("another", A) = 0, TF("another", B) = 1/5 = 0.2
    \item TF("text", A) = TF("text", B) = 1/5 = 0.2
    \item TF("document", A) = TF("document", B) = 2/5 = 0.4
\end{itemize}

\noindent\textbf{TF-IDF Vectors:}
\begin{itemize}
    \item Document A: $[0.06, 0, 0.06, 0.12]$
    \item Document B: $[0.06, 0.06, 0.06, 0.12]$
\end{itemize}

\section{Experimental Setup}\label{sec:experiment}

Our experimental setup aims at evaluating the effectiveness of NLP and machine learning techniques in malware classification. It consisted of the follows:

    \begin{itemize}
        \item \textbf{Host OS:} We used Ubuntu 18.04 LTS on a machine with an Intel i7 processor, 8GB RAM, and a 10TB HDD.  Cuckoo Sandbox 2.0.7 was employed for malware analysis on the Ubuntu host.

        \item We collected two lakhs diverse set of malware samples from VirusShare representing seven categories.
 
        \item A separate Windows 10 environment was used with an Intel i7 processor, 128GB RAM, and 5TB storage to collect and analyze the dynamic analysis reports obtained from Cuckoo Sandbox.
        \item Unigram, bigram and trigram analysis was conducted on Windows 10 host to transform API call sequences into $n$-grams of each categories.
    \end{itemize}

This setup allowed controlled experimentation to assess the potential of $n$-gram and machine learning techniques in enhancing the performance of malware classification.

Our proposed approach was implemented using Python 3.10.9 programming language. Comprehensive information about our malware detector's experimental setup, our dataset, and source code can be accessed on GitHub$^4$.



\section{Results Analysis}\label{sec:rana}
\begin{table}[ht]
\caption{Datasets used}
\begin{center}
 \begin{tabular}{|C{0.7cm}|C{2cm}|C{2cm}|C{2cm}|C{2cm}|}

\hline
\textbf{S.No} & \textbf{{Types}}& \textbf{{Test Sample}}& \textbf{{Train Sample}}& \textbf{{Total Sample}} \\
\hline\hline
1& Adware &406 & 1580 &\textbf{1986} \\
\hline
2& Backdoor &123 & 551  &\textbf{674}\\
\hline
3& Downloader &495 & 2002 &\textbf{2497}\\
\hline
4& Spyware &190 & 756 &\textbf{946}\\
\hline
5& Trojan &695 & 2873 &\textbf{3568}\\
\hline
6& Worm &277 & 1080 &\textbf{1357}\\
\hline
7& Virus &500 & 1892 &\textbf{2392}\\
\hline
8& Benign &1724 & 6910 &\textbf{8634}\\
\hline\hline
&\multicolumn{1}{|c|}{\textbf{Total}} & \textbf{4410} &\textbf{17644} &\textbf{22054}\\
\cline{2-5} 
\hline
\end{tabular}
\label{tab:mal_data}
\end{center}
\end{table}
\noindent The dataset in Table \ref{tab:mal_data} comprises of 22,054 malware samples categorized into 7 types (Adware, Backdoor, Downloader, Spyware, Trojan, Worm, Virus), alongside a Benign class. Training and testing samples are split in an 80:20 ratio per malware type, totaling 17,644 training samples and 4,410 testing samples. The Benign class has the most samples. This distribution ensures a balanced representation, facilitating thorough training and evaluation of machine learning models.
\noindent The dataset for this research shown in Table \ref{tab:mal_data} consists of malware samples categorized into 7 types: Adware, Backdoor, Downloader, Spyware, Trojan, Worm, and Virus, along with a class for Benign samples. The dataset is divided into training and testing samples, with varying ratios 80:20 for each malware type. The total number of samples in the dataset is 22,054, with 17,644 samples used for training and 4,410 samples used for testing. Each malware type has a different distribution of samples between the training and testing sets, with the Benign class having the highest number of samples. The dataset's distribution suggests a relatively balanced representation of malware types, allowing for comprehensive training and evaluation of machine learning models.

\begin{figure}[hbt!]
    \centering
       \begin{subfigure}[b]{0.49\textwidth}
        \centering
        \includegraphics[width=\textwidth]{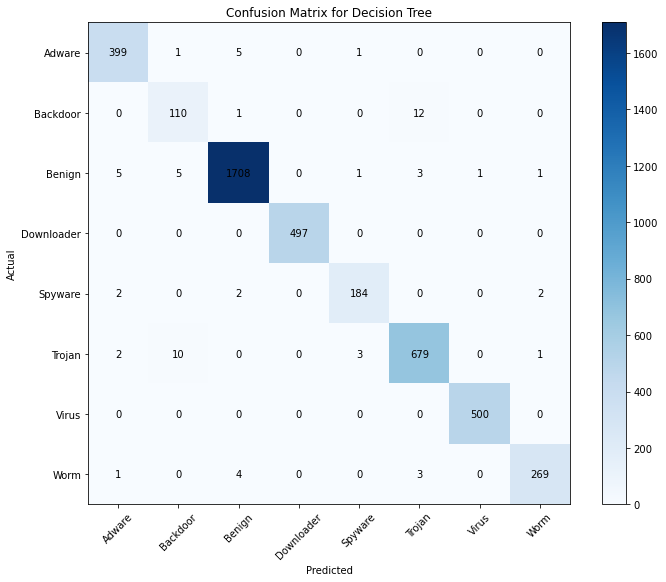}
        \caption{Decision Tree}
        \label{fig:decision_tree}
    \end{subfigure}
    \begin{subfigure}[b]{0.49\textwidth}
        \centering
        \includegraphics[width=\textwidth]{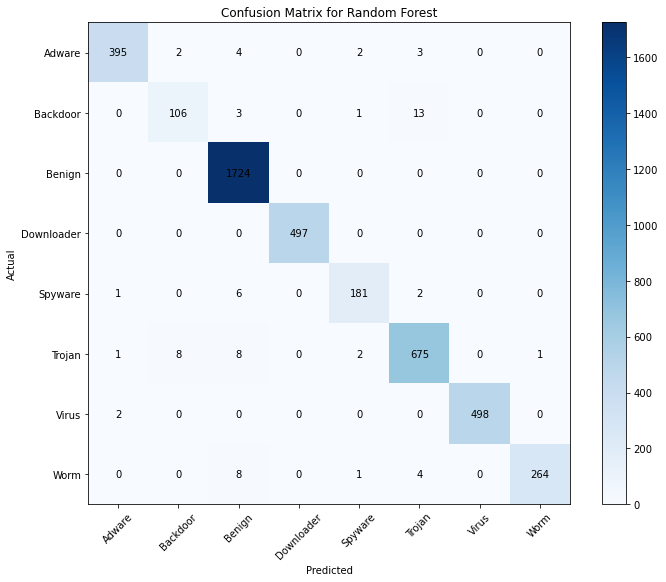}
        \caption{Random Forest}
        \label{fig:random_forest}
    \end{subfigure}
\end{figure}
 \begin{figure}[h!]
    \ContinuedFloat
    \centering
    \begin{subfigure}{0.49\textwidth}
        \includegraphics[width=\linewidth]{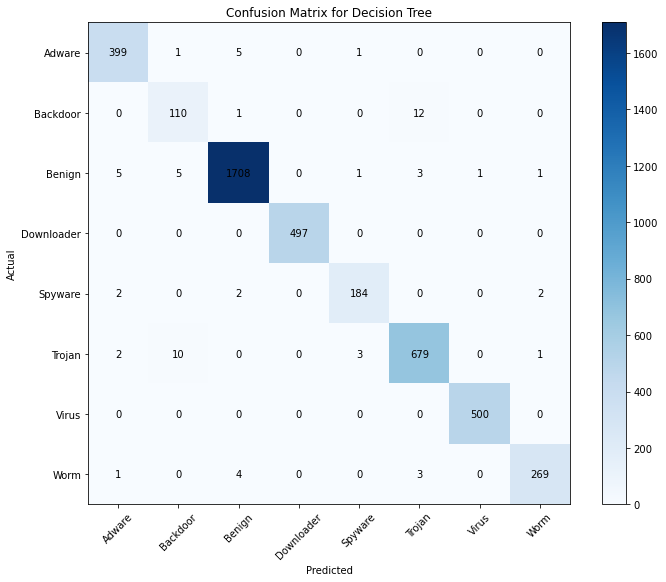}
        \caption{k-Nearest Neighbors}
        \label{fig:knn}
    \end{subfigure}
    \begin{subfigure}[b]{0.49\textwidth}
        \centering
        \includegraphics[width=\textwidth]{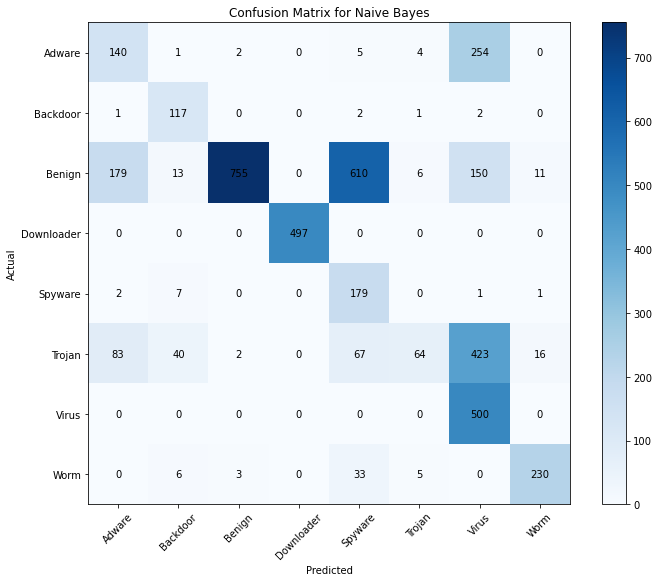}
        \caption{Naive Bayes}
        \label{fig:naive_bayes}
    \end{subfigure}
    \begin{subfigure}{0.49\textwidth}
        \includegraphics[width=\linewidth]{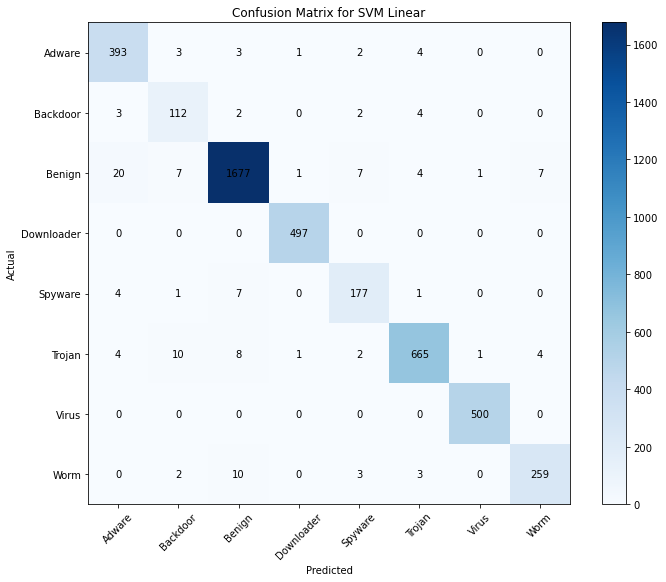}
        \caption{SVM Linear}
        \label{fig:svml}
    \end{subfigure}
        \begin{subfigure}{0.49\textwidth}
        \includegraphics[width=\linewidth]{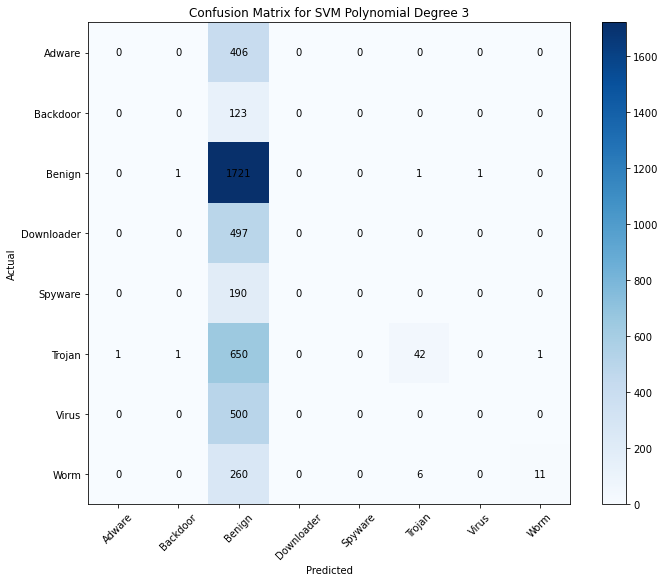}
        \caption{SVM Polynomial 3$^\circ$}
        \label{fig:svmpoly3}
    \end{subfigure}
     \begin{subfigure}{0.49\textwidth}
        \includegraphics[width=\linewidth]{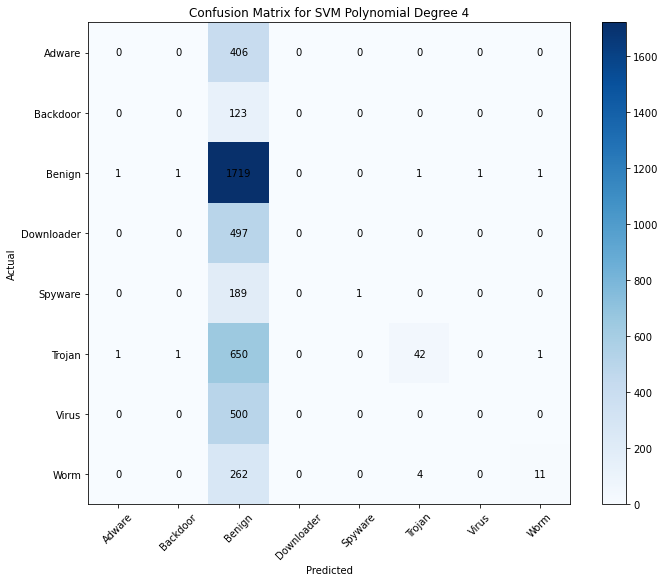}
        \caption{SVM Polynomial 4$^\circ$}
        \label{fig:svmpoly4}
    \end{subfigure}
     \begin{subfigure}{0.49\textwidth}
        \includegraphics[width=\linewidth]{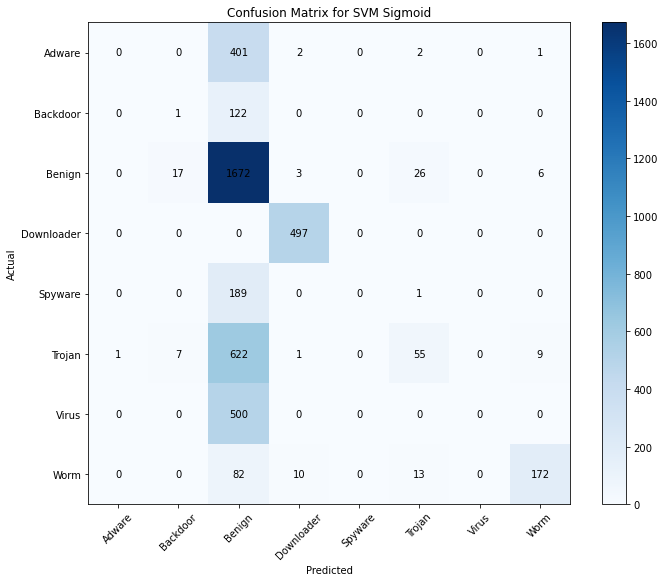}
        \caption{SVM Sigmoid}
        \label{fig:svms}
    \end{subfigure}   
\end{figure}
 \begin{figure}[h!]
    \ContinuedFloat
    \centering

    \begin{subfigure}{0.49\textwidth}
        \includegraphics[width=\linewidth]{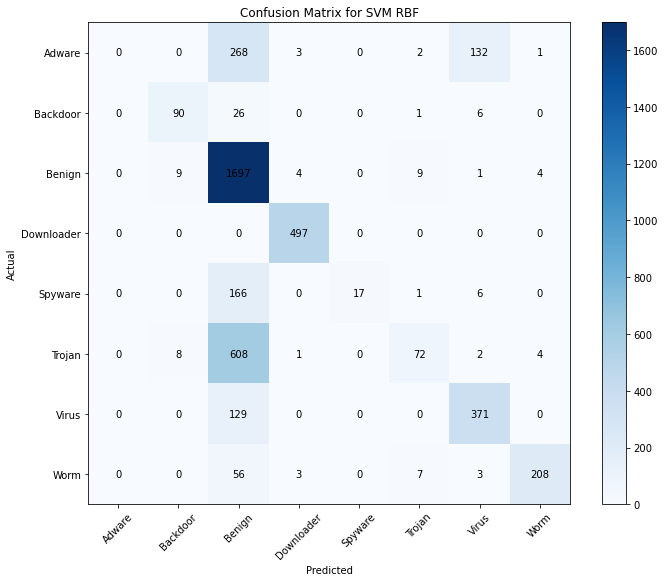}
        \caption{SVM RBF}
        \label{fig:svmrbf}
    \end{subfigure}
         \begin{subfigure}{0.49\textwidth}
        \includegraphics[width=\linewidth]{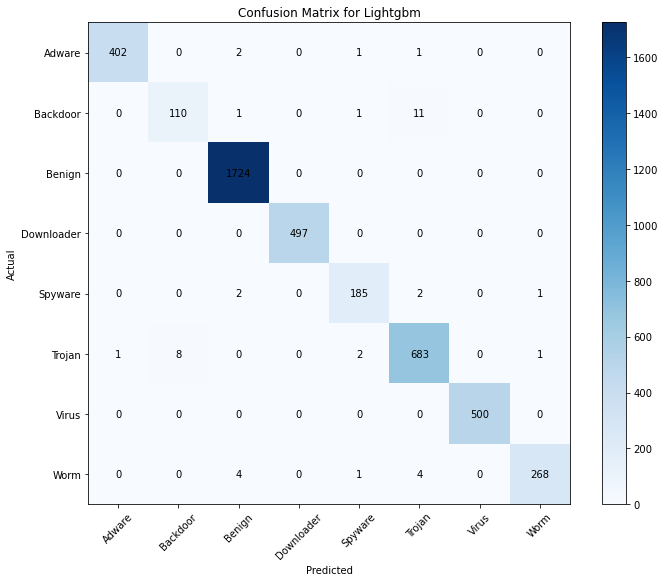}
        \caption{LightGBM}
        \label{fig:lightgbm}
    \end{subfigure}
    \begin{subfigure}{0.49\textwidth}
        \includegraphics[width=\linewidth]{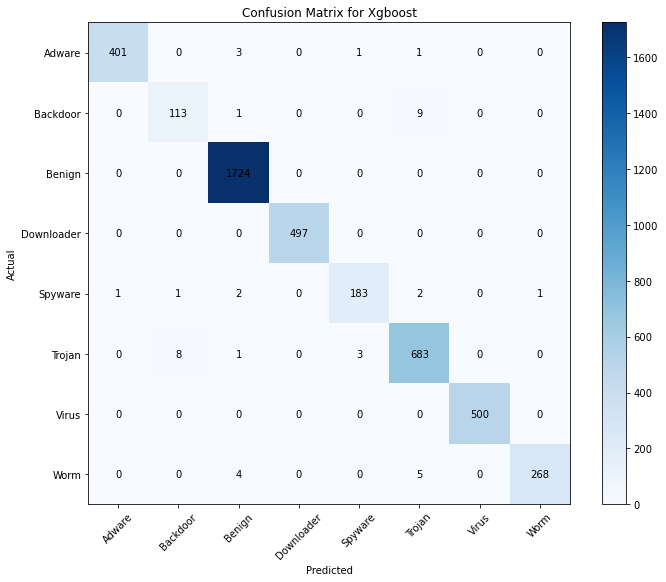}
        \caption{XGBoost}
        \label{fig:xgboost}
    \end{subfigure}
    \caption{Confusion Matrices for Malware Classification Using ML Techniques}
     \label{fig:conffig}
\end{figure}

Based on the evaluation metrics in Table \ref{tab:evaluationml}, it can be observed that the XGBoost classifier stands out as the best performer, achieving the highest accuracy, precision, recall, and $F_1$ score among all classifiers (99.02\%, 98.35\%, 97.74\%, and 98.04\%, respectively). Conversely, SVM Polynomial (degree 3 and 4) and SVM Sigmoid perform poorly, with accuracy rates around 40\% and very low precision, recall, and $F_1$ scores, making them unsuitable for this task.

\begin{table}[ht!]
\caption{Evaluation Metrics for malware classification using ML Techniques}
\begin{center}
\begin{tabular}{|C{.7cm}|C{3cm}|C{1.5cm}|C{1.5cm}|C{1.5cm}|C{1.5cm}|}
\hline
\multicolumn{1}{|c|}{} &\multicolumn{1}{|c|}{} &\multicolumn{4}{|c|}{\textbf{Performance Parameters}} \\
\hline \hline
\textbf{S.No} & \textbf{{Classifiers}}& \textbf{{Acc}}& \textbf{{$F_1$}} & \textbf{{Rec}}& \textbf{{Pre}} \\
\hline
1 &Decision Tree & 98.50 & 97.16 & 97.30 & 97.23 \\
\hline
2 &k-Nearest Neighbors & 94.40 & 93.16 & 92.16 & 92.65  \\
\hline
3 &Naive Bayes & 56.26 & 65.49 & 69.98 & 57.70  \\
\hline
4 &SVM Linear & 97.01 & 94.77 & 95.93 & 95.32 \\
\hline
5 &SVM Polynomial $3^{\circ} $ & 40.21 & 27.12 & 13.73 & 9.45  \\
\hline
6 &SVM Polynomial $4^{\circ} $ & 40.19 & 39.19 & 13.78 & 9.57  \\
\hline
7 &SVM RBF & 66.91 & 73.10 & 55.03 & 55.17 \\
\hline
8 &SVM Sigmoid & 54.33 & 36.96 & 33.48 & 31.32 \\
\hline
9 &Random Forest & 98.37 & 97.75 & 96.35 & 97.03  \\
\hline
10 &\textbf{XGBoost } &\textbf{99.02 } & \textbf{98.35} & \textbf{97.74} &\textbf{98.04 } \\
\hline
11 &LightGBM & \textbf{99.02} & 98.31 & 97.60 & 97.95  \\
\hline
\end{tabular}
\label{tab:evaluationml}
\end{center}
\end{table}

The poor performance of SVM Polynomial (degree 3 and 4) suggests potential overfitting due to complexity, while SVM Sigmoid struggles with capturing relationships in multi-class problems like malware classification. Naive Bayes exhibits lower performance due to underfitting from its simplicity. In contrast, XGBoost and LightGBM perform best, followed by Random Forest, Decision Tree, and SVM Linear, each demonstrating strengths in handling complex datasets and capturing important patterns. KNN performs decently but lags behind due to limitations in capturing complex data patterns as Confusion Matrices shown in Figure~\ref{fig:conffig}.
\section{Comparison of our work with present state-of-the-art techniques}\label{sec:compare}
Our work focuses on improving malware classification using NLP-based $n$-gram analysis. Since we utilize a unique dataset, we lack direct comparisons with existing state-of-the-art techniques. Our approach harnesses the power of Natural Language Processing (NLP) and $n$-gram analysis, offering a distinctive perspective on malware detection. In the absence of any directly related work, we compare our work with some closely related work

Dabas et al.\cite{dabas2023effective} achieved 99.6\% accuracy in Windows malware detection using TF-IDF enriched API calls, reducing the feature set to 9\%, but they used 2500 malware and 2500 benign samples.
Dabas et al.\cite{dabas2023malanalyser} achieved up to 99.7\% accuracy with 30\% features and 100\% accuracy on GA-enriched features, but they used 2500 malware and 2500 benign samples.
Sharma et al.\cite{sharma2022windows} achieved up to 99.91\% accuracy for SVM and Logistic Regression algorithms by combining API call feature sets into an integrated set and applying TF-IDF, but they didn't disclose the percentage of features used.
The detailed comparison of our work with some available related work is given in Table \ref{tab:com}.
\begin{table*}[hbt!]
\caption{Quantitative comparison of some malware detection techniques}
\begin{center}
\begin{tabular}{ |c|c|c|c|c|c|c|c|c|c| }
 \hline
 \textbf{No} &\textbf{Author} &\textbf{$n$-gram} &\textbf{IF-TDF} &\textbf{Feat. Red$^{n}$} &\textbf{ML Tech} &\textbf{Datasets} &\textbf{Class$^{n}$} &\textbf{Dect$^{n}$} &\textbf{Acc}\\
 \hline
  \hline
1 &Dabas et al.\cite{dabas2023effective}               & \ding{55}            & \ding{51}            & \textbf{9\%}             & \ding{51} &VirusShare &\ding{55} &\ding{51} & 99.6         \\
 \hline
2 &Dabas et al. \cite{dabas2023malanalyser}                & \ding{55}              & \ding{51}            & \textbf{30\% }             & \ding{51}         &VirusShare &\ding{55} &\ding{51} & 99.7    \\
 \hline
 3 &Sharma et al. \cite{sharma2022windows}                & \ding{55}              & \ding{51}            & \textbf{n.d.}             & \ding{51}        &VirusShare &\ding{55} &\ding{51} &  99.9    \\
 \hline

4 &Proposed Work                 & \ding{51}              & \ding{51}            & \textbf{1.6\% }              & \ding{51}      & VirusShare&\ding{51} &\ding{55} &  99.02      \\
\hline
\end{tabular}
\label{tab:com}
\end{center}
\end{table*}
\section{Conclusions and Future Work}\label{sec:futurework}
In conclusion, the evaluation of machine learning algorithms identify XGBoost and LightGBM classifiers as the most proficient options among those considered. Both classifiers demonstrate exceptional performance across key metrics, achieving the highest accuracy, precision, recall and  $F_1$ score values. XGBoost achieves remarkable results with an accuracy of 99.02\%, precision of 98.35\%, recall of 97.74\%, and $F_1$ score of 98.04\%, closely followed by LightGBM with comparable scores of 99.02\% accuracy, 98.31\% precision, 97.60\% recall, and 97.95\% $F_1$ score. These findings show the effectiveness and reliability of both XGBoost and LightGBM for classification tasks, making them the optimal choices for achieving high predictive performance and robustness in similar scenarios.

We outline the following possible extensions to enhance our work: 1) Exploring diverse feature selection approaches to enhance analysis versatility. 2) Applying various deep learning approaches to broaden the research scope and uncover new insights. 3) Utilizing genetic algorithm approaches for feature selection and creation to further expand potential insights from our investigation.


\bibliographystyle{unsrt}
\bibliography{references}


\end{document}